\documentclass[
    ,final            
  ]
  {aipproc}

\layoutstyle{6x9}

\def\gtrsim{\ifmmode {\mathbin{\lower 3pt\hbox   
    {$\,\rlap{\raise 5pt\hbox{$\char'076$}}\mathchar"7218\,$}}}
    \else {${\mathbin{\lower 3pt\hbox
    {$\rlap{\raise 5pt\hbox{$\char'076$}}\mathchar"7218\,$}}}
    $}\fi}
\def\lesssim{\ifmmode {\,\mathbin{\lower 3pt\hbox   
    {$\,\rlap{\raise 5pt\hbox{$\char'074$}}\mathchar"7218\,$}}}
    \else {${\mathbin{\lower 3pt\hbox
    {$\rlap{\raise 5pt\hbox{$\char'074$}}\mathchar"7218\,$}}}
    $}\fi}

\begin{document}

\title{The Final Parsec Problem}

\author{Milo\v s Milosavljevi\' c$^1$ \& David Merritt$^2$}
{
  address=
  {
    $^1$ Theoretical Astrophysics, California Institute of Technology,
    Pasadena CA 91125; milos@tapir.caltech.edu \\
    $^2$ Department of Physics and Astronomy, Rutgers University, 
    New Brunswick, NJ 08903; merritt@physics.rutgers.edu
  }
}

\begin{abstract}
The coalescence of massive black hole binaries is one of the main sources of
low-frequency gravitational radiation that can be detected by LISA.  When two
galaxies containing massive black holes merge, a binary forms at the center of
the new galaxy.  We discuss the evolution of the binary after its separation
decreases below one parsec.  Whether or not stellar dynamical processes can
drive the black holes to coalesce depends on the supply of
stars that scatter against the binary.  We discuss various mechanisms by which
this supply can be replenished after the loss cone has been depleted.
\end{abstract}

\maketitle

\vspace*{-13.1cm}
{\footnotesize{
\begin{verbatim}
review, to appear in: `The Astrophysics of Gravitational Wave Sources',
J. Centrella (ed), AIP, in press   (2003) 
\end{verbatim}
}}
\vspace*{10.1cm}


\section{Introduction}

The prospect that low-frequency gravitational radiation will be detected
by
LISA has recently energized theoretical inquiries into the formation and
the evolution of massive black hole binaries (MBHB).  MBHB are sources of low-frequency gravitational radiation that will provide highest signal/noise for LISA, but the event rate for these sources is much less well known than for the other two principal astrophysical sources, compact binaries and extreme mass ratio inspiral (see contributions by G.~Nelemans and S.~Sigurdsson, this volume).  Astronomical
evidence
for the existence of black holes with masses $M_{\rm bh}\gtrsim10^6
M_\odot$ in
galaxy spheroids with central stellar velocity dispersions
$\sigma\gtrsim100\textrm{ km s}^{-1}$ is increasingly compelling; the
evidence
for black holes with masses $100-10^6 M_\odot$ in low-dispersion
spheroids is
still equivocal. When two galaxies merge, a MBHB forms at the center of
the new galaxy \cite{Begelman:80,Roos:81}.  
There has been considerable interest in determining whether 
black hole coalescence occurs efficiently following 
galaxy mergers,
since almost all predictions of MBH coalescence event rates 
equate the galaxy merger rate -- derived from 
models of structure formation in which galaxies merge hierarchically \cite{Hae94,MHN01,VHM03,WL03,JB03} -- with the MBH coalescence rate.  
Detailed estimates of the MBH coalescence rate are fundamentally limited by the resolution of electromagnetic telescopes.  It is therefore expected that MBH coalescences as observed by LISA will facilitate the first definite conclusions regarding galaxy merger rates and MBH demographics at high redshifts 
\cite{Hughes:02}. 

The mutual gravitational capture of the black holes 
is facilitated by the dynamical drag imparted to
the orbiting MBHs by the overlapping galaxies.  
The inner parts of elliptical galaxies and many spiral bulges are 
well described by power-law
(``cuspy'') stellar density profiles of the form $\rho\sim r^{-\gamma}$
with
$\gamma\approx 2$ \cite{geb96}.  
High luminosity ellipticals often exhibit a
shallowing
of the density profile ($\gamma<2$) near the very centers, where the
MBHs are
located.  Numerical simulations of galaxy mergers have shown that the
black
holes remain embedded in their donor cusps throughout the merger 
\cite{Milosavljevic:01}.
Since the dynamical drag is a function of the combined black hole and
stellar
cusp mass, which exceeds that of the black hole 
until the very final stages of the merger,
the black hole inspiral and the MBHB formation take place on a
dynamical time
scale for black holes of comparable mass.  In the unequal mass case, the
infall time scale is lengthened proportionally to the black hole mass
ratio $M_1/M_2$ ($M_1\geq M_2$), but for $M_2\gtrsim 10^6 M_\odot$
remains short compared to time scale for the subsequent 
evolution \cite{Merritt:00}, which is the subject of this contribution.  

\section{Gravitational Slingshot and Binary Decay}

The galaxy merger delivers the black holes to within the distance
$a_{\rm
  hard}=G\mu/4\sigma^2$, where $\mu=M_1M_2/(M_1+M_2)$ is the reduced
mass.
  This distance is about 1 parsec for $M_1=M_2=2\times10^7M_\odot$.  At
that point, the binary continues to decay by scattering stars
super-elastically \cite{sva74}.
 Stars in the merged galaxy with orbits approaching the binary closely
  enough to be perturbed by the rotating quadrupole component of the
binary's
  potential belong to the ``loss cone.'' The loss cone is defined in
analogy
  with a similar structure characterizing the distribution of the
stellar-mass
  objects around solitary massive black holes \cite{Frank:76,baw76}.
When a star inside the loss
  cone impinges on the MBHB, it exchanges kinetic energy with the binary
  and is shot out at an average velocity comparable to the binary's
orbital
  velocity $\bar v_*\sim \sqrt{G(M_1+M_2)/a}\gg\sigma$.  This is a form of
the
  gravitational slingshot mechanism commonly used to accelerate
spacecraft in
  the solar system.
As a result, the binary's binding energy increases and its semi-major
axis $a$
  decreases.  

In a crude approximation, the factor by which the binary
  separation decays can be related to the total mass in stars $M_{\rm
scat}$ that are scattered against the binary between times $t_1$ and
$t_2$ via \cite{Quinlan:96}:
\begin{equation}
\label{eq:semievol}
a(t_2)\sim a(t_1)\exp \left\{-\frac{M_{\rm scat}(t_1,t_2)}{{\cal
J}(M_1+M_2)}\right\} ,
\end{equation}
where ${\cal J}$ is the mass ejection coefficient which in fact depends weakly on $a/a_{\rm hard}$ and the black hole masses.  

Approximation (\ref{eq:semievol}) is widely used in the semi-analytic modeling of the MBHB population dynamics within hierarchical structure formation scenarios.  Value of the mass ejection coefficient has been estimated from three-body scattering experiments \cite{Quinlan:96}:
\begin{equation}
{\cal J}\ \approx\  0.1\ \ln\left(\frac{4 a_{\rm hard}}{a}\right) \ \approx \ 0.1-1 .
\end{equation}
This agrees with the value ${\cal J}\approx 0.5$ measured in $N$-body simulations of equal-mass MBH mergers \cite{Milosavljevic:01}.  If the galaxy is nearly spherical and the ejected stars return to interact with the binary more than once, the efficiency of mass ejection depends on the potential well depth $\Delta\Phi$ separating the energies at which stars enter and exit the loss cone \cite{Milosavljevic:03}:
 \begin{equation}
{\cal J}\approx \frac{1}{(\Delta\Phi/2\sigma^2)(a/a_{\rm hard})} .
\end{equation}
To shrink the binary by one $e$-folding, a stellar mass of ${\cal J}M_{\rm bh}$ must be transported from an energy marginally bound to the black hole, to the galactic escape velocity.

\section{Coalescence}
If the semi-major axis decreases until it becomes less than \cite{Peters:64}:
\begin{equation}
a_{\rm gr} = 
\left\{\frac{64}{5}\frac{G^3M_1M_2(M_1+M_2)F(e)}{c^5} 
t_{\rm gr}\right\}^{1/4} ,
\end{equation}  
the emission of gravity waves leads to the coalescence of the black
holes on
a time scale $t_{\rm gr}$.  Here, $F(e)$ is an eccentricity-dependent
factor
equal to unity for a circular binary. $N$-body simulations of MBHB
formation
in galaxy mergers suggest that the eccentricities remain moderate, 
although our understanding of the MBHB eccentricity evolution is still incomplete.
The factor by which the binary must shrink from its conception to 
coalescence is:
\begin{equation}
\frac{a_{\rm hard}}{a_{\rm gr}}
\approx 34 \times e^{9.2/\alpha}
\left(\frac{M_1+M_2}{10^6M_\odot}\right)^{1/4-2/\alpha}
p^{3/4}(1+p)^{-3/2} \left(\frac{t_{\rm gr}}{10^9\textrm{
yrs}}\right)^{-1/4} ,
\end{equation}
where $p=M_2/M_1\leq 1$ and $\alpha=d\log M_{\rm bh}/d\log \sigma\approx4-5$
is the logarithmic slope of the tight relation between the black hole
mass and the central velocity dispersion of the galaxy 
\cite{fem00,geb00}.

Therefore, to achieve coalescence in a Hubble time, 
the MBHB must shrink by $4-5$ $e$-foldings, which requires
the scattering of $10-20\times\mu$ worth of stars from the loss cone.   
After its initial emptying
and the accompanying rapid shrinking of the binary by a factor of
$\sim5$ past
$a_{\rm hard}$ (as observed in $N$-body simulations with $\rho\sim
r^{-2}$
initial density profiles \cite{Milosavljevic:01}), 
the mass of the loss cone $\lesssim\mu$ if
the
galaxy is approximately spherical.  Once this mass is expended, the
binary is
still a factor $\gtrsim 10$ wider than the separation at which it would
coalesce gravitationally in a Hubble time; the binary decay may
therefore
stall \cite{Roos:81,Milosavljevic:01,val96,gor00,Yu:02a}.  
The problem is exacerbated if the initial cusp profile is shallower than
$r^{-2}$.
The apparent inability of stellar dynamical processes to drive the
binaries to coalescence poses a potential problem for the detection of
these sources by LISA.  
Circumstantial evidence, however, suggests that
long-lived MBHBs may be rare: no
smoking-gun detections of MBHBs have been reported 
(with the possible exception of OJ 287; see \cite{Pursimo:00}),
although a number of AGN have time-dependent features
that have been provisionally attributed to MBHBs
(Komossa, this volume).
The theoretical difficulty of shrinking a MBHB by a factor of $\sim 100$
after its formation at a separation of $\sim 1$ pc is
called the ``final parsec problem.''

\section{Loss-Cone Diffusion}
The final parsec problem is most severe in nearly spherical galaxies
where
the mass inside the loss cone is the smallest.  The loss cone boundary
is
defined by the minimum angular momentum $J_{\rm
loss}\sim\sqrt{G(M_1+M_2)a}$
that a star can have and still avoid being perturbed by the MBHB.  
Consider 
the state of the loss cone after all the stars inside have been
scattered {\it once}.  
Since the depletion of the orbital population inside 
the loss cone leads to a stalling in the decay rate, continued decay
hinges on 
the rate at which stars
diffuse into the loss cone via collisional relaxation.  The relaxation
can be
modeled by means of the orbit-averaged Fokker-Planck equation \cite{lis77} 
describing
the
evolution of the phase space density $f(E,J,t)$ subject to the boundary
condition $f(E,J_{\rm loss},t)=0$; i.e., stars are assumed to be 
removed from the system by the gravitational slingshot
once they straddle the loss cone boundary.  
The loss cone flux is proportional to
the gradient of the phase space distribution at its boundary, ${\cal
F}\propto
\partial f/\partial J|_{J_{\rm loss}}$.  

Although it is tempting
to seek a steady-state solution 
$f_{\rm equi}(E,J)\propto \log(J/J_{\rm loss})$ \cite{cok78,mat99,Yu:02a},
in reality the center of the galaxy is not likely to be collisionally relaxed 
\cite{fab97}, and thus 
the distribution of stars near the loss cone is never in a steady state
on time scales of order the relaxation time.  Indeed, the final stages
of a
galaxy merger when the MBHB forms proceed in a time much shorter than
the
galaxy crossing time.  
Therefore the distribution function $f(E,J)$ immediately following 
the formation of a hard binary can be far from that describing
a steady-state flux of stars into the loss cone.
Sudden draining of the loss cone during formation of the hard binary
produces steep phase space gradients 
that are closer to the step function:
\begin{equation}
f(E,J) \approx   
\cases{ \bar f(E), & $J>J_{\rm loss}$ \cr
  0 , & $J<J_{\rm loss}$ } .
\end{equation}
Since the collisional transport rate in phase space is proportional
to the gradient of $f$ with respect to $J$,
steep gradients imply an enhanced flux into the loss cone.
The depletion of stars {\it outside} the loss cone affects the
density profile of the galaxy and is identified 
with the cusp destruction. 
The broken power-law
profiles of high-luminosity elliptical galaxies can be interpreted as
fossil
evidence for this process \cite{Milosavljevic:02,Ravindranath:02}.  

\begin{figure}
  \resizebox{\hsize}{!}{
    \includegraphics{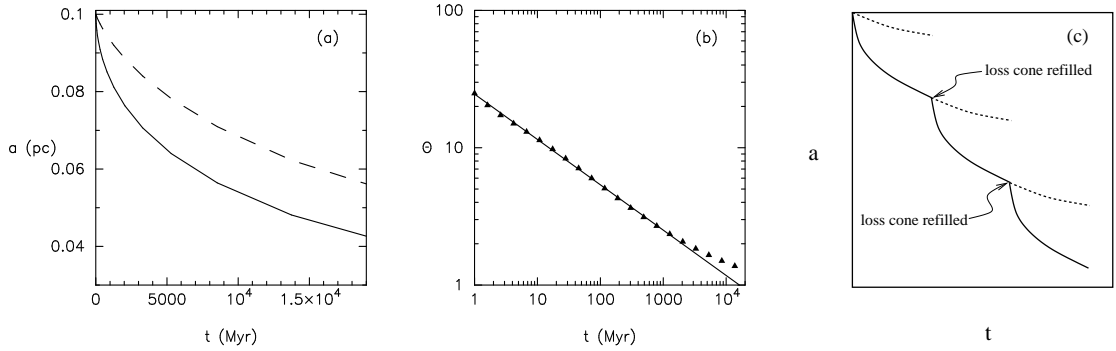}}
  \caption{(a) Evolution of the semi-major axis $a$ in a merger of equal 
    galaxies with
    $\sim10^6 M_\odot$ black holes starting from the initial
    separation of $a(0)=0.1$ pc (solid line).
    The evolution is always more rapid than
    that predicted assuming that the galaxy is
    collisionally relaxed (dashed line).\label{fig:noneq} (b)
Enhancement of
    the decayed compared with that of the collisionally-relaxed, 
    steady-state solution, expressed in terms of $\Theta=\Delta
    a/\Delta a_{\rm equi}$ (triangles) and a power-law fit (solid line).
(c)
    Schematic illustration of the evolution of the semi-major axis in
the
    presence of episodic refilling of the loss cone. 
    (From \cite{Milosavljevic:03}.)
    }
\end{figure}

The time evolution of the stellar
distribution near the loss cone is an initial value problem equivalent
to
the diffusion of heat in cylindrical coordinates \cite{Frank:76}.  
Ignoring the diffusion in
$E$, the Fokker-Planck equation for diffusion in $J$ 
reads ($J\geq J_{\rm loss}$): 
\begin{equation}
\label{eq:fp}
\frac{\partial f(E,J,t)}{\partial t} = 
\frac{\lambda(E)}{J} \frac{\partial}{\partial J}
\left[J \frac{\partial f(E,J,t)}{\partial J}\right] ,
\end{equation}
where $\lambda(E)$ is related to the orbit-averaged diffusion
coefficient.  
Since the boundary angular
momentum $J_{\rm loss}$ decreases with time, equation (\ref{eq:fp}) can be solved
iteratively by discretizing the decrements in $J_{\rm loss}$ and
interpolating
$f$ between these via the Fourier-Bessel synthesis \cite{Milosavljevic:03}.  
Solutions obtained this way
can be compared to the collisionally relaxed, 
steady state ($\partial f/\partial t=0$) 
solution normalized to the isotropic distribution (Figure
\ref{fig:noneq}a).  In an example scaled to the galaxy M32 with a
$3\times10^6M_\odot$ black hole, 
the binary in the exact solution has decayed only
$\sim30\%$ more than that in the steady-state solution.  The difference
between the two, however, is much more substantial early on (Figure
\ref{fig:noneq}b), which is of crucial importance if episodic, violent
dynamical perturbations such as the infall of satellite galaxies or
giant
molecular clouds rejuvenate the loss cone \cite{Zhao:02} 
by restoring the steep phase-space
gradients instrumental for the enhanced diffusion (Figure
\ref{fig:noneq}c).   
For example, if the episodic replenishment in a galaxy like M32 occurs
every 
10, 100, or 1,000 Myrs, the average MBHB decay rate will be
$\Theta\approx$ 
10, 5, or 3 times higher than what
the equilibrium theory would have implied \cite{Milosavljevic:03}.  

Brownian motion of the MBHB in
the neighborhood of the geometric center of the galaxy
can to some extent mimic the effects of collisional relaxation and drive
stars
into the loss cone.  The Brownian motion results from the equipartition
of
kinetic energy between the MBHB and the stars in the galaxy, $\langle
v_{\rm
  brown}^2\rangle\sim (m_*/M_{\rm bh}) \sigma^2$, where $m_*$ is the
average
stellar mass \cite{mer01,Chatterjee:02,Dorband:03}.  As the binary
wanders in space, it can sweep up stars that would have remained just
outside
the loss cone for a static binary.  The time scale on which the loss
cone
refills in this fashion is:
\begin{equation}
t_{\rm brown}\sim 400\textrm{ Myr} \times
\frac{a}{a_{\rm hard}}
\left(\frac{m_*}{M_\odot}\right)^{-1}
\left(\frac{M_1+M_2}{10^6M_\odot}\right)^{2-3/\alpha} e^{14/\alpha} 
K(E) ,
\end{equation}
where $K(E)$ is a function of the orbital 
energy such that $K(2\sigma^2)\sim 1$.  
The amplitude of the Brownian motion is only modestly enhanced
by ``super-elastic scattering'' by the binary \cite{mer01}.
In galaxies $t_{\rm brown}\gtrsim 1$
Gyr and thus the Brownian motion probably fails to significantly enhance
the flux into the loss cone, 
in spite of earlier suggestions to the contrary \cite{Quinlan:97}.  
In $N$-body simulations, however, 
the Brownian motion {\it saturates} the loss
cone flux which is one of the many pitfalls that plague the numerical
modeling of MBHBs.

\section{Spherical and Aspherical Galaxies}

Mode of interaction between the stars in a galaxy and a MBHB is 
also influenced by the 
geometry of the galactic potential, which could either be nearly spherical
or substantially aspherical (axisymmetric, triaxial, or irregular).

In spherical galaxies all stars that are candidates for slingshot ejection by the MBHB encounter the binary within one crossing time from the moment the binary forms.  
The time-dependent loss cone solution derived in the previous section
was based on
the ``sink'' paradigm, in which a star is lost from the system
as soon as it transgresses the loss cone boundary.
This model is valid in the case of capture or tidal disruption of
stars by a single black hole but is less relevant to MBHBs, 
since stars that interact with the binary simply receive kicks 
($\Delta E, \Delta J$) that transport them to another orbit without 
necessarily ejecting them from the galaxy.
If the binary orbit decays on a time scale longer than the orbital
period of an interacting star, stars inside the loss cone can
remain inside the loss cone after ejection, encountering the binary
again at their next pericenter passage.
In principle a star can interact many times with the binary
before being ejected from the galaxy or falling outside the loss
cone; each interaction takes additional energy from the binary
and hastens its decay.

\begin{figure}
  \resizebox{\hsize}{!}{
    \includegraphics{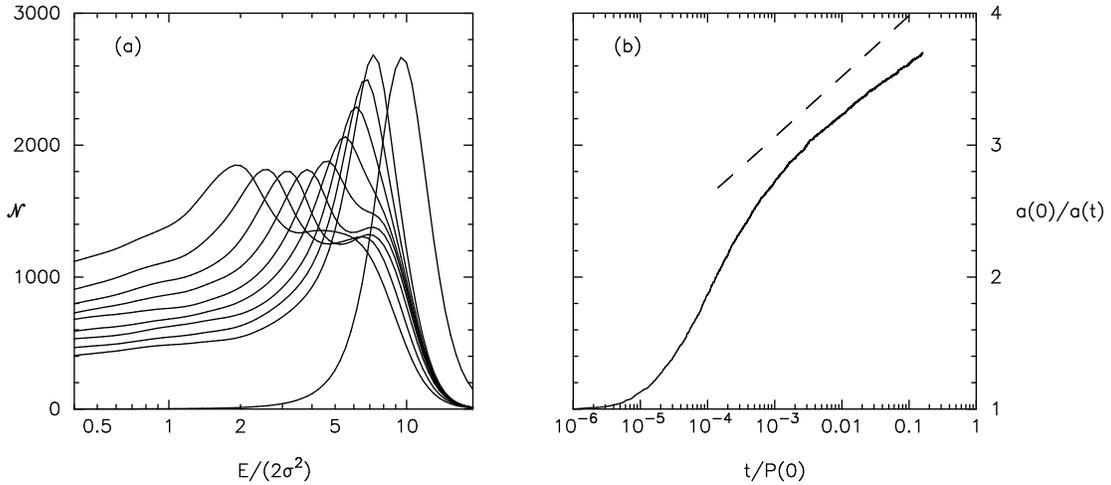}}
  \caption{
(a) Snapshots of the binding energy distribution ${\cal N}(E,t)$ of the
stars residing inside the loss cone at $t=0$.  From right to left the data
were taken at exponentially-increasing time intervals.  As the binary
separation decreases and the velocity of the slingshot ejection increases,
stars inside the loss cone are heated through repeated scattering and
shift to smaller $E$; a significant fraction of them remain inside the
loss cone at all times.  (b) Evolution of the semi-major axis exhibiting
the logarithmic behavior motivated in the text (solid line).  The slope of
$da/d\ln t$ is close to that given by the theoretical prediction
(\ref{eq:reeject}); $P(0)$ is the dynamical time of the potential well in
which the galaxy is embedded (dashed line).  (From \cite{Milosavljevic:03}.)
    \label{fig:reeject}
    }
\end{figure} 

We illustrate this ``secondary slingshot'' with a simple model
in the spherical geometry.
Consider a group of ${\cal N}$ stars of mass $m_*$ and energy per unit mass
$E$ that interact with the binary and receive a mean energy increment 
of $\langle\Delta E\rangle$.
Averaged over a single orbital period $P(E)$, the binary hardens
at a rate \cite{Milosavljevic:03}:
\begin{equation}
{d\over dt}\left({GM_1M_2\over 2a}\right) = m_*{\langle\Delta  
E\rangle {\cal N} \over P(E)}.
\label{eq:edot}
\end{equation}
In subsequent interactions, the number of stars that remain
inside the loss cone scales as $J_{\rm loss}^2\propto a$, while
the ejection energy is $\sim \Phi_{\rm eject} + G(M_1+M_2)/2a\sim a^{-1}$.
Hence ${\cal N}\langle\Delta E\rangle\propto a^1a^{-1}\propto a^0$.
Assuming a singular isothermal sphere for the galaxy potential, we
derive the result \cite{Milosavljevic:03}:
\begin{eqnarray}
\label{eq:reeject}
\frac{a(0)}{a(t)}&\sim&
1+\frac{4\sigma^2 a(0)}{G(M_1+M_2)}
\ln\left[1+\frac{m_*{\cal N}\langle\Delta E\rangle}{2\mu\sigma^2}
  \frac{t}{P(E_0)}\right]\nonumber\\
&\sim&1+0.25\ \ln \left[\textrm{few}\times 10 \times
\frac{t}{P(E_0)}\right] .
\end{eqnarray}
Hence the binary's binding energy increases as the logarithm of the
time.
Figure \ref{fig:reeject} illustrates the evolution of $a^{-1}$ in an $N$-body
simulation
where star-star interactions have been replaced by a smooth potential
to inhibit relaxation.
The observed rate of decay $da^{-1}/d\ln t\approx 1.7\times 10^4$
is close to the prediction of equation (\ref{eq:reeject}), where we have
$4\sigma^2/G(M_1+M_2)=2\times 10^4$.
Re-ejection at the rate of equation (\ref{eq:reeject}) would by itself
induce changes in $a^{-1}$ by factors of a few over a Hubble
time, in addition to the shrinkage due to collisional loss cone
refilling.  For $P(E_0)=10^{3,5,7}$ years, we obtain $a(0)/a(t_{\rm Hubble})\approx
5$, $4$, and $3$, respectively.

In non-spherical galaxies where 
the total angular momentum $J$ is not a conserved quantity, 
there exists a potentially larger population of orbits
that can encounter the binary but only once per several orbital
periods.
The mechanisms of loss cone evolution and re-ejection discussed
above in the spherical geometry would be modified somewhat in
axisymmetric \cite{Yu:02a} or triaxial galaxies \cite{Merritt:03}.
The triaxial case is potentially the most interesting:
stellar bars are commonly observed in galactic nuclei, 
and torques from barlike potentials are often invoked
to drive gas inflows during the quasar epoch
\cite{Shlosman:90}.
Since orbital angular momentum is not conserved
in the triaxial geometry, a large fraction of the stars
in a triaxial bar can potentially interact with the central 
MBHB.
These ``centrophilic'' orbits are typically chaotic
due to scattering off the central mass;
in spite of their unfavorable time-averaged shapes,
chaotic orbits can make up 50\% or more of the total mass
of a triaxial nucleus \cite{Poon:02}.

Numerical integrations \cite{Merritt:03}
reveal that the frequency of pericenter passages with $r_{\rm peri}<a$
for a chaotic orbit of energy $E$ is
roughly linear in $a$, ${\cal N}(r_{\rm peri}<a)\approx A(E)a$, 
up to a maximum pericenter distance of $r_{\rm peri,max}(E)$.
The total rate at which stars pass within a distance $a$ of
the center is therefore:
\begin{equation}
\dot M \approx a\int A(E){\cal M}_{\rm chaos}(E)dE
\end{equation}
where ${\cal M}_{\rm chaos}(E)dE$ is the mass on chaotic orbits in the energy
range $E$ to $E+dE$.
In a triaxial nucleus with density
$\rho\sim r^{-2}$ and central mass $M_{\rm bh}$,
the numerical integrations reveal:
\begin{equation}
A(E) \approx 5{\sigma^5\over
G^2M_{\rm bh}^2}e^{-\left(E-\Phi(r_{\rm bh})\right)/\sigma^2},\ \ \ 
r\gtrsim r_{\rm bh}\equiv{GM_{\rm bh}\over 2\sigma^2}.
\end{equation}
The feeding rate due to stars with energies $E>\Phi(r_{\rm bh})$ is then
\begin{equation}
\dot M \approx 
{\cal F}_{\rm chaos} {\sigma^3\over G}{a\over r_{\rm bh}}\approx 10^3M_\odot\
{\rm yr}^{-1} {{\cal F}_{\rm chaos}\over 0.5} \left({\sigma\over 200\ {\rm km\
s}^{-1}}\right)^3 {a\over r_{\rm bh}}
\end{equation}
with ${\cal F}_{\rm chaos}$ the fraction of stars on chaotic orbits.
Even a small ${\cal F}_{\rm chaos}$ implies a substantial rate of supply to
a MBHB when it first forms, with $a\sim a_{\rm hard}\sim
(\mu/2M_{\rm bh})r_{\rm bh}$.
Such high feeding rates would imply substantial changes in
MBHB separations even if triaxial distortions to the potential
were transient, due for instance to mergers or accretion events.

\section{$N$-Body Simulations}

Several attempts have been made to model the formation 
and the long-term evolution of MBHBs using large-scale
$N$-body simulations \cite{mak97,Quinlan:97,Milosavljevic:01,Hemsendorf:02}. 
Because of the discreteness effects associated 
with approximating a galaxy with $N\gtrsim 10^9$ 
stars by a model consisting of only, at best, $N\lesssim 10^6$ particles, 
the applicability of direct $N$-body simulations appears to be 
limited to the early stage of the MBHB evolution.  
The rapidity with which a galactic merger proceeds guarantees
that the discreteness effects do not affect the state 
of the galaxy just after the merger is complete.  
That point marks the transition 
to a much more gradual decay in the MBH separation 
when the effective relaxation 
time in the simulation can easily become shorter than the decay time.

Simulations fail to reproduce the 
long-term evolution correctly because in the simulations the loss cone is
almost completely full, while in real galaxies it is largely empty.
In the general case there exists a critical stellar orbital 
energy $E_{\rm crit}$ separating the full region $E<E_{\rm crit}$ 
(large radii)
from the empty region $E>E_{\rm crit}$ (small radii).
Assuming a density profile $\rho\sim r^{-2}$ and  
a potential of the form $2\sigma^2\ln(r/r_0)$ such that 
$r_0=10^3GM_{\rm bh}/2\sigma^2$ we find \cite{Milosavljevic:03}:
\begin{equation}
\label{eq:ecrit}
E_{\rm crit}\approx 2\sigma^2 \ln \left(\frac{7.5\times10^4}{{\cal N}_{\rm bh}} 
\frac{GM_{\rm bh}/8\sigma^2}{a}\right).
\end{equation}
where ${\cal N}_{\rm bh}=M_{\rm bh}/m_*$ is the 
number of stars that make the mass of the black hole.
The transition from a full to an empty loss cone happens when $E_{\rm crit}$ becomes smaller than a few $\times 2\sigma^2$, implying that
${\cal N}_{\rm bh}\sim 10^4-10^5$.  Since a typical MBH contains
0.1\% of its host galaxy's mass \cite{mef01a}, and thus $N\sim
10^3{\cal N}_{\rm bh}$, an $N$-body simulation would have to contain
$10^{4-5}\times10^3=10^{7-8}$ bodies to reproduce the correct,
diffusive behavior of a real galaxy.  
This requirement is a severe one for direct-summation $N$-body codes,
which rarely exceed particle numbers of $\sim 10^6$ even on
parallel hardware \cite{Dorband:03}.
One route might be to couple the special purpose GRAPE
hardware\footnote{http://grape.astron.s.u-tokyo.ac.jp/grape/},
which is limited to $N\lesssim 10^6$, to algorithms that
can handle large particle numbers by swapping with a fast front end.

\section{Outstanding Problems}
We conclude by mentioning two outstanding problems related to 
the dynamics of MBHB binaries.
Although the gravitational back-reaction tends to circularize 
binary black holes, some residual eccentricity can remain, 
especially if stellar dynamical processes prior to the 
emission of gravitational waves act to amplify the eccentricity to large values.  Residual eccentricities result in the excitation of higher harmonics in the signal \cite{Peters:63}, thereby complicating the detection of a gravitational wave event severely.  
In spite of the importance of eccentricity for detection, an accurate and general evaluation of the MBHB eccentricity evolution remains a challenge.

Similarly, large-mass-ratio black hole binaries deserve extra attention.  Although the best understood MBHBs are those of consisting of nearly equal-mass black holes, it is probable that most coalescences involve black holes of very unequal mass. For example, intermediate-mass black holes (IMBHs) with masses of $10^{2-4}M_\odot$ may be able to form in young star clusters---such the Arches and the Quintuplet clusters in the Milky Way bulge \citep{Fig02,FMM99}---via the segregation of massive stars to the cluster center \citep{Spi69}, followed by the runaway growth in stellar mergers and the collapse of the agglomerate star into an IMBH \citep{PZM02,RFG03}. An IMBH makes its way to the galaxy center and forms a binary with the nuclear MBH \citep{HM03}.  Therefore, large-mass-ratio MBHBs are expected to exist even in galaxies that had not experienced recent mergers.  The orbital evolution of these systems remains another challenge to dynamical exploration.  

\section{Conclusions}
We have focused on stellar dynamical mechanisms for extracting
energy from a MBHB, but other schemes are possible and even likely.
We note a close parallel between the ``final parsec problem''
and the problem of quasar fueling: in both cases, a quantity
of mass of order $\sim 10^8M_\odot$ must be supplied to the
inner parsec in a time much shorter than the age of the universe.
Nature clearly accomplishes this in the case of the quasars,
probably through gas flows driven by torques from stellar bars 
\cite{Shlosman:90}.
The same inflow of gas could contribute to the decay of a MBHB,
by leading to renewed star formation or rapid accretion of gas
\cite{Armitage:02}.  Similarly, the presence of a third 
black hole in a galactic nucleus could accelerate the
decay by increasing 
the MBHB's eccentricity through the Kozai mechanism \cite{Blaes:02}, 
or by extracting the
binary's energy via the triple black hole slingshot interaction 
\cite{Valtonen:94}.

%

\begin{theacknowledgments}
This work was supported by NSF grants AST 00-71099 
and by NASA grants NAG5-6037 and NAG5-9046 to DM, and by a Sherman
Fairchild Postdoctoral Fellowship to MM.
\end{theacknowledgments}

\end{document}